%% file: Manuscript.tex
\documentclass{article}

\usepackage{microtype}
\usepackage{graphicx}
\usepackage{subcaption}
\usepackage{booktabs} 
\usepackage{multirow}
\usepackage{tikz}
\usetikzlibrary{calc}
\usepackage{balance}

\usepackage{hyperref}

\usepackage[accepted]{icml2023}

\usepackage{amsmath}
\usepackage{amssymb}
\usepackage{mathtools}
\usepackage{amsthm}

\usepackage[capitalize,noabbrev]{cleveref}

\theoremstyle{plain}
\newtheorem{theorem}{Theorem}[section]

\theoremstyle{definition}

\theoremstyle{remark}
\newtheorem{remark}[theorem]{Remark}

\usepackage[textsize=tiny]{todonotes}

\icmltitlerunning{ }

\input{Our_macros.tex}

\begin{document}

\twocolumn[
\icmltitle{A Ridgelet Approach to Poisson Denoising}




\begin{icmlauthorlist}
\icmlauthor{Ali Dadras}{yyy}
\icmlauthor{Klara Leffler}{yyy}
\icmlauthor{Jun Yu}{yyy}
\end{icmlauthorlist}

\icmlaffiliation{yyy}{Department of Mathematics and Mathematical Statistics, Ume{\aa} University, Ume{\aa}, Sweden}

\icmlcorrespondingauthor{Ali Dadras}{ali.dadras@umu.se}
\icmlcorrespondingauthor{Klara Leffler}{klara.leffler@umu.se}

\icmlkeywords{ }

\vskip 0.3in
]



\printAffiliationsAndNotice{ }  

\begin{abstract}
This paper introduces a novel ridgelet transform-based method for Poisson image denoising. Our work focuses on harnessing the Poisson noise's unique non-additive and signal-dependent properties, distinguishing it from Gaussian noise. 
The core of our approach is a new thresholding scheme informed by theoretical insights into the ridgelet coefficients of Poisson-distributed images and adaptive thresholding guided by Stein's method.

We verify our theoretical model through numerical experiments and demonstrate the potential of ridgelet thresholding across assorted scenarios. Our findings represent a significant step in enhancing the understanding of Poisson noise and offer an effective denoising method for images corrupted with it. \\
\end{abstract}


\section{Introduction}\label{Introduction}
Image noise can be seen as a random variation within the image information. It is usually generated in the physical acquisition process and affects visual effects such as the image colour or brightness. Therefore, image denoising is a fundamental task in computer vision. The additive white Gaussian noise model is by far the most adopted. However, in imaging techniques where the acquisition process involves a counting process, the data are more accurately modelled by a Poisson distribution. Such processes occur in e.g. astronomy and astrophysics \citep{lanteri2005restoration}, spectral imaging \citep{priego20174dcaf}, and biomedical imaging \citep{thanh2015denoising}, including photon-counting imaging such as positron emission tomography and computed tomography \citep{vardi1985statistical, hasinoff2014photon, thanh2019review}. 

Assuming an image $\X=[X_{i,j}]$ follows a Poisson model, its pixel values are viewed as Poisson-distributed random variables. Hence, given a noiseless image $\Lambda$ with pixel values $\lambda_{i,j} \geq 0$ at position $(i,j)$, the probability of observing a noisy pixel value $X_{i,j}$ is given by $X_{i,j} \sim \mathrm{Po} (\lambda_{i,j})$ where
\begin{align}\label{eq:Poisson}
P(X_{i,j}= x) = \frac{\lambda_{i,j} ^{x} e^{-\lambda_{i,j}} }{x!}, 
\end{align}
where $(i, j) \in \mathbb{N}^N \times \mathbb{N}^M$. It is worth noting that Poisson noise is not additive and that its strength is directly dependent on the image intensity. From Equation \eqref{eq:Poisson}, one can conclude that the signal-to-noise ratio (SNR) in each pixel is $\sqrt{\lambda_{i,j}}$, implying that lower intensity in observed images comes with stronger noise \citep{giryes2014sparsity, zhang2019vst}. The mentioned SNR of the observed noisy Poisson image is proportional to the noiseless image \citep{zhang2021poisson}. 
Poisson noise substantially differs from Gaussian noise. 
With its direct dependency between signal and noise, it does not follow the simple additivity principle of white Gaussian noise. 
Furthermore, the variance is equal to the mean for Poisson-distributed variables, implying high variance for high-intensity levels. 
Therefore, traditional approaches for Gaussian noise removal do not directly apply to Poisson denoising \citep{thanh2019review, zhang2019vstnet}. 
Denoising of Poisson-distributed images aims at estimating the underlying noiseless intensity profile $\Lambda$ from the noisy images $\X$. Therefore, defining the noise power in the image by its peak value, i.e., the maximum value of the underlying intensity profile $\Lambda$ is natural. 

\subsection{Related Work}
Methods for Poisson denoising can roughly be divided into two categories: direct and indirect. 
The indirect approaches are based on variance stabilisation transforms (VST) that remove the signal-dependency property such that an additive Gaussian model can approximate the noise. A classical Gaussian denoising approach can then be applied before inversely transforming the denoised image back to the original domain. The Anscombe \citep{anscombe1948} or Haar-Fisz \citep{fryzlewicz2004haar} transforms are the most commonly used. An extension of the Anscombe transform was presented incorporating multiscale VST that can be combined with thresholding in the wavelet, ridgelet or curvelet domains \citep{zhang2008wavelets}. Combinations of VST and the popular Gaussian denoising technique of block-matching and 3D filtering have also been proposed for both high- and low-intensity data \citep{dabov2007image, azzari2016variance}.

The direct approaches aim to use the Poisson characteristics of the noise. Examples include total variation (TV) regularisation based on the gradient information of the image \citep{rudin1992tv}. The TV-based approach encourages piecewise smoothness, exploiting that images often exhibit sharp transitions only at edges, allowing for effective noise reduction while preserving edges. Extensions into the Bayesian setting have been proposed, incorporating Poisson statistics into the data fidelity term \citep{le2007variational, figueiredo2010admm}. To overcome unwanted staircase effects resulting from the TV approaches, higher-order TV denoising was proposed \citep{chan2000hotv}. Sparsity-based regularisation has also been proposed, aiming to find the sparsest representation of a noisy image in terms of linear combinations of a dictionary of basis elements and reconstruct a clean image from the sparse representation. Examples include low-rank methods combined with sparse coding \citep{dong2012lowrank} and patch-based sparse representations combined with dictionary learning \citep{giryes2014sparsity}. 
 
Multiresolution strategies to Poisson intensity estimation were introduced to address potential correlations amongst the elements of $\X$. Statistical modelling of transform coefficients as latent variables, e.g. via filterbanks such as wavelets, is convenient since such transforms tend to generate temporal and spatial decorrelated coefficients as well as energy compaction properties for a variety of data \citep{hirakawa2011skellam}. Furthermore, the Poisson statistics are known to be preserved across scales in the low-pass channels of an unnormalised Haar wavelet transform \citep{luisier2010purelet}. Examples include generalised wavelet approaches with scale-dependent thresholds \citep{kolaczyk1996wavelet, kolaczyk1999wavelet, charles2003wavelet} and multiscale Bayesian approaches \citep{timmermann1999multiscale, lu2004multiscale}. 
A Poisson unbiased risk estimator based on a linear expansion of thresholds (PURE-LET) was proposed following Stein's unbiased risk estimator (SURE) for the Gaussian noise model \citep{luisier2010purelet}. There, the thresholds are adapted to local estimates of the signal-dependent noise variance derived from the corresponding low-pass coefficients at each scale. The same idea was also presented based on Poisson risk estimation via the Skellam distribution in a Bayesian setting \citep{hirakawa2011skellam}. 
As an alternative to pure wavelets and to better capture anisotropic features sucha as lines, the Ridgelet transform was proposed \citep{candes1999ridgelet}, moving the wavelet analysis into the Radon domain. Implementations of the ridgelet transform are based on either the finite Radon transform \citep{do2000ridgelet} or radial discrete lines in the Fourier domain \citep{carre2004ridgelet}. Poisson denoising via ridgelet transform has been implemented via integration of dual-tree complex wavelet transform of the radon coefficients \citep{chen2007ridgelet}, geometric multiscale ridgelet support vector transform \citep{yang2013ridgelet}, and patchwise finite ridgelet transform \citep{liu2014ridgelet}. 

Apart from the model-based methods described above, recent developments in learning have motivated data-driven approaches to address the Poisson noise. 
Motivated by the traditional VST methods, three convolutional neural subnetworks were employed to simulate the forward transform, the Gaussian denoising, and the inverse transform via the so-called VST-Net \citep{zhang2019vstnet}. 
Direct learning strategies that focus on the Poisson characteristics have, for instance, taken inspiration from Gaussian image denoising, e.g. the trainable nonlinear reaction-diffusion model for Gaussian denoising of images \citep{chen2016reactiondiffusion}, where an extension was presented for fast and accurate Poisson denoising, replacing the reaction term by a Poisson noise-derived function \citep{feng2017diffusion}. 
A deep convolutional denoising network (DenoiseNet) was proposed for low-light images \citep{remez2017denoisenet}. An improvement to deep convolutional neural networks was also presented with additional multi-directional long-short-term memory networks to better capture and learn the statistics of the residual noise components \citep{kumwilaisak2020variational}. 
Recently, a variational Bayesian deep network was proposed for blind Poisson denoising, combining subnetworks for noise estimation and removal \citep{liang2023variational}.

\subsection{Contribution and Outline}
We aim to develop a novel apporach for Poisson denoising based on the ridgelet transform. In this paper we, therefore, investigate the distributional properties of the noisy image in its ridgelet domain and propose a proper thresholding scheme for noise removal.  
The manuscript is structureed as follows. In Section \ref{Theoretical Results} we derive the distribution of the ridgelet coefficients; and in Section \ref{Experiments} we perform numerical experiments to verify the theoretically derived distributions and illustrate the potential of ridgelet thresholding. 
We conclude in Section \ref{Discussion} with some discussion.

\bigskip
\section{Theory and Methods}\label{Theoretical Results}
In this section, we will investigate the distributional properties of the ridgelet coefficients of the Poisson-distributed image. 

We model the image data as 
\begin{equation}
    \X=  \T(\Lambda)
    \label{eq:noise_model}
\end{equation}
where $\X$ is a matrix of measurements, $\Lambda$ is a noiseless data matrix, and $\T(.)$ is a signal dependent Poisson matrix with $\mathbb{E}[\T(\Lambda)]= \Lambda$. \\

\bigskip
\subsection{Preliminaries}\label{Preliminaries}

The ridgelet transform was introduced in 1999 as a geometrical generalisation of wavelets, applying one-dimensional wavelet transforms to projections in the Radon domain \citep{candes1999ridgelet}. For this study to be self-contained, we include some preliminary theories about the Radon and ridgelet (wavelet) transforms.

\subsubsection{The Radon Transform}
The Radon Transform in its continuous form is a fundamental operation in image processing, e.g. computed tomography (CT) and positron emission tomography (PET). It involves integrating an image function $f(\s)$ along straight lines in the plane. Each line is uniquely defined by its perpendicular distance $r\in(-\infty,\infty)$ from the origin and the angle $\theta\in[0,\pi)$ the line makes with the horzontal axis. Mathematically, the Radon transform can be expressed as
\begin{align*}
 \mathbb{R}_f(r,\theta) = \int_{\mathbb{R}^2} f(\s)\delta(s_1 \cos\theta+ s_2\sin\theta -r) d\s
\end{align*}
where $\delta$ represents the Dirac delta function. By varying $r$ and $\theta$, the Radon transform captures the global structure of $f(\s)$ from different perspectives. 
Since the Radon transform transforms a point $\delta$-function in $f$ into a sinusoidal line $\delta$-function in $\mathbb{R}_f$, the transform function is often referred to as a sinogram \citep{Press2006}. 
The inverse Radon transform is utilised in PET imaging to reconstruct cross-sectional images from sinogram projections. 

\subsubsection{The Ridgelet Transform}
The ridgelet transform is built on combining the Radon and wavelet transforms. 
The (separable) continuous wavelet transform of a signal $f(\s)\in L^2(\mathbb{R})$ at scales $\mathbf{a}$ and locations $\mathbf{b}$ is defined by 
\begin{align*}
W(\mathbf{a}, \mathbf{b})=\int_{\mathbb{R}^2} f(\s)\psi_{\mathbf{a}, \mathbf{b}}(\s) d\s,
\end{align*}
where the wavelet functions in two dimensions are tensor products
\begin{align*}
\psi_{\mathbf{a}, \mathbf{b}}(\s) =  \psi_{a_1, b_1}(s_1) \psi_{a_2, b_2}(s_2)
\end{align*}
of one-dimensional scaled and shifted wavelet functions $\psi_{a,b}(t)=a^{-1/2}\psi((t-b)/a)$, for $a>0$ and $b\in\mathbb{R}$.

Given its construction, the basic strategy for calculating the continuous ridgelet transform is first to compute the Radon transform and then apply a one-dimensional wavelet transform to the Radon projections $\R_f(\theta,\cdot)$, i.e.
\begin{align*}
R(a, b,\theta) =\int_{\mathbb{R}} \R_f (r,\theta) \psi_{a, b}(r)  dr.
\end{align*}

\bigskip
\subsection{Distribution of wavelet coefficients}
\label{subseq:Distribution of wavelet coefficients}
When operating the different transforms on the image data, their discrete versions are applied. 

The discrete wavelet transform of a vector $\x \in \mathbb{N}^N$ is based on the discretised version of a wavelet function $\psi$ at scale $j$ and location $k$. The wavelet coefficient $W_{j,k}$ of $\x=(x_1, ..., x_N)$ at  $(j,k)$ is then the projection of $\x$ onto a wavelet $\psi_{j,k}$, given as an inner product  \citep{jensen2001ripples} by 
\begin{equation}
\label{eq:Discrete wavelet}
W_{j,k}= \sum_{i} \psi_{j,k}[i]* x_i.
\end{equation}
\begin{theorem}[Distribution of wavelet coefficients]
\label{theorem:Distribution of wavelet coefficients}
Let $\x \in \mathbb{N}^N$ be a vector of independent Poisson random variables $x_i$ with intensities $\lambda_i$.
 The distribution of the discrete wavelet coefficients $W_{j,k}$ of $\x$ can be approximated by the distribution $\mathcal{S}(\Tilde{\lambda}_{j,k}^+,\alpha_{j,k}^+,\Tilde{\lambda}_{j,k}^-, \alpha_{j,k}^-)$ that is the distribution of the scaled Poisson differences $W_{j,k}= \alpha_{j,k}^+ \ \Tilde{W}_{j,k}^+ - \alpha_{j,k}^- \ \Tilde{W}_{j,k}^-$ where $\Tilde{W}_{j,k}^\pm$ are Poisson variables with intensities $\Tilde{\lambda}_{j,k}^\pm$ and 
\begin{align*}
\Tilde{\lambda}_{j,k}^\pm  &= \frac{\Big(\sum_{i \in \mathcal{I}_{j,k}^\pm} \lambda_i \ {\psi_{j,k,i}}\Big)^2}{\sum_{i \in \mathcal{I}_{j,k}^\pm} \lambda_i \ {(\psi_{j,k,i})}^2}  
    \nonumber \\
\alpha_{j,k}^\pm &= \frac{\sum_{i \in \mathcal{I}_{j,k}^\pm} \lambda_i \ {(\psi_{j,k,i})}^2}{\sum_{i \in \mathcal{I}_{j,k}^\pm} \lambda_i \ {|\psi_{j,k,i}}|}.  
\end{align*}
where $W_{j,k}= \sum_{i} \psi_{j,k,i} x_i$  as defined in \cref{eq:Discrete wavelet}  and we define 
$\mathcal{I}_{j,k}^+ = \{i| \psi_{j,k,i} \geq 0\}$ and $\mathcal{I}_{j,k}^- = \{i| \psi_{j,k,i} < 0\}$.

\noindent {\bf Proof.}
\cm{ The discrete wavelet transform of $\x$ is defined as 
 \begin{equation}
     \W= \Tilde{W} \x
 \end{equation}
 where $\Tilde{W} \in \R^{n \times n}$ denotes the forward wavelet transform matrix.}
We begin by splitting 
\cref{eq:Discrete wavelet} into two parts
 \begin{equation*} \label{Distribution of wavelet coefficients:eq:wavelet coef def}
     W_{j,k}= W_{j,k}^+ - W_{j,k}^-,
 \end{equation*}
 where
 \begin{align*}
 W_{j,k}^+= \sum_{i \in \mathcal{I}_{j,k}^+} \psi_{j,k,i} x_i
 \\
 W_{j,k}^- =\sum_{i \in \mathcal{I}_{j,k}^-} |\psi_{j,k,i}| x_i,
 \end{align*}
where for any $i$ in $\mathcal{I}_{j,k}^+$, $\psi_{j,k,i}$ is positive, and for any $i$ in $\mathcal{I}_{j,k}^-$ is negative.

Note that  $W_{j,k}^+$ and $W_{j,k}^-$ are
the sum of weighted Poisson variables.
Let us compute the mean and variance of $W_{j,k}^+$ and $W_{j,k}^-$
\begin{align*}
\mu_{W_{j,k}^+}&= \sum_{i\in \mathcal{I}_{j,k}^+} \lambda_i \ \psi_{j,k,i}
, \quad \mu_{W_{j,k}^-}&= \sum_{i\in \mathcal{I}_{j,k}^-} \lambda_i \ |\psi_{j,k,i}|
\nonumber \\
\sigma^2_{W_{j,k}^+} &= \sum_{i\in \mathcal{I}_{j,k}^+} \lambda_i \ \psi_{j,k,i}^2,
\quad 
\sigma^2_{W_{j,k}^-} &= \sum_{i\in \mathcal{I}_{j,k}^-} \lambda_i \ \psi_{j,k,i}^2
\end{align*}
Using the properties of scaled Poisson distribution (SPD) \cite{bohm2014statistics}, we can approximate the distributions with the distribution of scaled Poisson random variables $\hat{W}_{j,k}^+:= \alpha_{j,k}^+ \ \Tilde{W}_{j,k}^+$ and 
$\hat{W}_{j,k}^-:= \alpha_{j,k}^- \ \Tilde{W}_{j,k}^-$
where $\Tilde{W}_{j,k}^+ \sim \mathrm{Po}(\Tilde{\lambda}^+_{j,k})$ and 
$\Tilde{W}_{j,k}^- \sim \mathrm{Po}(\Tilde{\lambda}^-_{j,k})$
. We use moment matching to find the best parameters for $\hat{W}_{j,k}$
\begin{align*}
\mu_{\hat{W}_{j,k}^+}&=  \alpha_{j,k}^+ \Tilde{\lambda}_{j,k}^+ = \sum_{i \in \mathcal{I}_{j,k}^+}\lambda_i \ {\psi_{j,k,i}}
\nonumber \\
\mu_{\hat{W}_{j,k}^-}&=  \alpha_{j,k}^- \Tilde{\lambda}_{j,k}^- = \sum_{i \in \mathcal{I}_{j,k}^-} \lambda_i \ {|\psi_{j,k,i}|}
\nonumber \\
\sigma^2_{\hat{W}_{j,k}^+} &=  \big(\alpha_{j,k}^+\big)^2 \Tilde{\lambda}_{j,k}^+ = \sum_{i \in \mathcal{I}_{j,k}^+} \lambda_i \ \psi_{j,k,i}^2
\nonumber \\
\sigma^2_{\hat{W}_{j,k}^-} &=  \big(\alpha_{j,k}^-\big)^2 \Tilde{\lambda}_{j,k}^- = \sum_{i \in \mathcal{I}_{j,k}^-} \lambda_i \ \psi_{j,k,i}^2.
\end{align*}
This will give us the parameters of SPDs.
\begin{align*}
\Tilde{\lambda}_{j,k}^\pm  &= \frac{\Big(\sum_{i \in \mathcal{I}_{j,k}^\pm} \lambda_i \ {\psi_{j,k,i}}\Big)^2}{\sum_{i \in \mathcal{I}_{j,k}^\pm} \lambda_i \ {(\psi_{j,k,i})}^2}  
    \nonumber \\
\alpha_{j,k}^\pm &= \frac{\sum_{i \in \mathcal{I}_{j,k}^\pm} \lambda_i \ {(\psi_{j,k,i})}^2}{\sum_{i \in \mathcal{I}_{j,k}^\pm} \lambda_i \ {|\psi_{j,k,i}}|}.  
\end{align*}
This completes the proof.  {\hfill $\square$}
\end{theorem}

\bigskip
\subsection{Discrete Radon transform}\label{subseq:Discrete Radon transform}
In this sub-section, we would like to calculate the distribution of the discrete Radon transform. Dealing with discrete data, there is a vast number of discrete versions of the Radon transform (DRT).

\bigskip
\subsubsection{Discrete Approximations of Lines}
In order to compute the discrete Radon transform, we need to define the discrete approximations of lines. Here one can use different line definitions such as  G\"{o}tz-Druckm\"{u}ller-Brady approach or rotaion based line definition to approximate lines \citep{gotz1996fast,brady1998fast}.
The DRT algorithm sums an image's pixel values along a
set of aptly chosen discrete lines, complete in slope and intercept. Different line approximations come with different computational complexities and approximation errors. 
 For an image of size $N \times N$, G\"{o}tz-Druckm\"{u}ller-Brady algorithm has a computational complexity of $\mathcal{O}(N^2 \log (N))$. This method approximates a continuous line with intercept $h$ and
slope $s/(N-1)$. One can show that its maximum vertical deviation from
the continuous line is less than or equal to $\frac{1}{6} \log_2 (N)$ \citep{brady1998fast}. 
 
Let $\X=[X_{i,j}]$ be an image matrix of size $N$, 
we deﬁne a set of lines $\DN(h, s)$ that transect the image, passing exactly through one array point in each column of the array and parameterised by integers $h$ and $s$. 

\cm{
The line $\DN(h, s)$ connects the entry $X_{0,h}$ to $X_{N-1,h+s}$. 

We focus on lines with slopes between $0^\circ$ and ${45}^\circ$ associated with $0 \leq s \leq N-1$. These discrete lines can be defined recursively as follows
\begin{align*}
\DN (h, 2s)&= \DNN(h, s) \ \cup \ \DNN(h+s, s)
\nonumber \\
\DN(h, 2s+1)&=\DNN(h, s) \ \cup \ \DNN(h+s+1, s)
\end{align*}
where $\cup$ joins the left and right halves.} 

\cm{
\begin{remark}[The number of points in $\DN (h, s)$]

By noting the fact that $s$ defines the slope of $\DN (h, s)$, one can count the number of points in the line $\DN (h, s)$ for the fixed $s$ as
\begin{equation*}
    |\DN (h, s)|= N (1+\frac{s}{N-1}).
\end{equation*}
\end{remark}
}

 Equipped with these discrete line definitions, we can define the DRT as follows
 \begin{equation*}
     \Ra(h, s) = \sum_{(i, j) \in \DN(h, s)}   \Xij,
 \end{equation*}
 where we assume $\Xij= 0$ if $i$ or $j$ is outside the range $[0, N-1]$.

 \cm{
 The superscript $(1)$ is used to indicate the “quadrant” $0^\circ$ to ${45}^\circ$. For other angle ranges, we compute DRT by appropriately ﬂipping the original image, then repeating exactly the same algorithm as described above for $\Raa(h,s)$.
\begin{align*}
\Rab(h, s) &= \sum_{(i, j) \in \DN(h, s)}   \Xij    \quad \text{(${45}^\circ$ to ${90}^\circ$)}
\\
\Rac(h, s) &= \sum_{(i, j) \in \DN(h, s)}   \XiNj \quad \text{(${-90}^\circ$ to ${-45}^\circ$)}
\\
\Rad(h, s) &= \sum_{(i, j) \in \DN(h, s)}   \XNji\quad \text{(${-45}^\circ$ to ${0}^\circ$)}.
\end{align*}
In each quadrant, the intercept $h$ varies in the vertical direction, while the rise $s$ varies horizontally.}

\cm{
\begin{remark}
DRT always sums one value in each of $N$ columns, independent of the angle, meaning that $\Ra^{(q)}(h, s)$ is exactly the sum of $N$ entries of the matrix $\X$. In other words $|\DN(h,s)|= N$.   
\end{remark}
}

\bigskip
\subsubsection{Distribution of Radon transform}

Using the DRT defined in the previous part, we can calculate the distribution of the DRT.

\bigskip
\begin{theorem}[Distribution of the DRT]
\label{theorem:Distribution of the DRT}
Let $\X$ be a random matrix with Poisson distributed entries with intensities $\lambda_{i,j}$ for each entry. The discrete Radon transform $\Ra(h, s)$ of $\X$ have Poisson distribution with the following intensity
\begin{align*}
\lambdahs &= \sum_{(i, j) \in \DN(h, s)}   \lambdaij 
\end{align*}

\noindent {\bf Proof.}  It is clear that $\Ra(h, s)$ is the sum of Poisson random variables with intensity $\lambdaij$ where $\{ (i, j) \in \DN(h, s) \}$, therefore, it is Poisson distributed with intensity $\lambdahs:= \sum_{(i, j) \in \DN(h, s)} \lambdaij$. 

\cm{Using the same reasoning, for the other quadrants we have
\begin{align*}
\lambdabhs &= \sum_{(i, j) \in \DN(h, s)}   \lambdaji 
\\
\lambdachs &= \sum_{(i, j) \in \DN(h, s)}   \lambdaiNj 
\\
\lambdadhs&= \sum_{(i, j) \in \DN(h, s)}   \lambdaNji. 
\end{align*}}
{\hfill $\square$}

\end{theorem}

\bigskip
\subsection{Distribution of Ridgelets}
Combining the result from \cref{subseq:Discrete Radon transform} and \cref{subseq:Distribution of wavelet coefficients}, we can derive the distribution of the Ridgelet coefficients. Ridgelets are defined as the wavelet transform of the Radon transform of the data with fixed slops, say $s$.

\bigskip
\begin{theorem}[Distribution of Ridgelet coefficients]
Let $\X$ be a random matrix with independent Poisson distributed entries with intensities $\lambda_{i,j}$ for each entry. The distribution of the discrete Rigelet transform $\Ri_{\X}$ of $\X$ can be approximated by the difference of the scaled Poisson random variables as in \cref{theorem:Distribution of wavelet coefficients}.

\noindent {\bf Proof.}
Combining the results of \cref{theorem:Distribution of wavelet coefficients} and \cref{theorem:Distribution of the DRT}, we get the result. {\hfill $\square$}
\end{theorem}

\bigskip
\subsection{Thresholding using Stein's method}
\label{subseq:Thresholding using Stein method}
 We denote the distribution of the difference of the scaled Poisson variables
 $W_{j,k}= \alpha_{j,k}^+ \ \Tilde{W}_{j,k}^+ - \alpha_{j,k}^- \ \Tilde{W}_{j,k}^-$
 with $\Sdist$ and the distribution of the addition of the scaled Poisson variables
 $W_{j,k}= \alpha_{j,k}^+ \ \Tilde{W}_{j,k}^+ - \alpha_{j,k}^+ \ \Tilde{W}_{j,k}^-$ with $\Qdist$. Note that $\Tilde{W}_{j,k}^+ \sim \mathrm{Po}(\Tilde{\lambda}^+_{j,k})$ and 
 $\Tilde{W}_{j,k}^- \sim \mathrm{Po}(\Tilde{\lambda}^-_{j,k})$.
  Let us assume that $w \sim \Sdist$ and $v \sim \Qdist$. Considering the estimator $\hat{\Lambda}(w,v)= w+ G(w,v)$ for the vector $[\Tilde{\lambda}^{+}_{j,k}, \Tilde{\lambda}^{-}_{j,k}]$, One can use Stein's method to find the optimal threshold $\tau$ for soft thresholding  $\hat{\Lambda}(w;\tau)=\mathrm{sgn}(w) \cdot \max\{ |w|-\tau, 0 \}$ \citep{hirakawa2011skellam}.
  
  \cm{
  $\ell^2$ risk of $\hat{\Lambda}(w,v)$ can be written as

 \begin{align}
 \mathbb{E}  \Big[ \nsq{\hat{\Lambda}(w,v)- [\Tilde{\lambda}^{+}_{j,k}-\Tilde{\lambda}^{-}_{j,k}]  }\Big] = ....
 \end{align}
 where $...$ as an unbiased estimator of the parameters.
}

\bigskip
\section{Experiments}\label{Experiments}
\bigskip
\subsection{Verification of Noise Distribution}\label{sec:Verification of Noise Distribution}
In this sub-section, we will present numerical experiments to verify the theoretically derived distributions of the Radon and ridgelet coeeficients of Poisson-distributed images. 

The assumed signal-dependent noise model is given by the relationship  $\X= \T(\Lambda)$ in Equation (\ref{eq:noise_model}). 
For the homogeneous case, we consider a constant-intensity 2D image corrupted with Poisson noise. We generate 1000 samples of the transforms and evaluate the disributional proerties within transform coefficients. Baring the Poisson assumption in mind, we estimate the mean and variance of the transform coefficients and report on mean-variance ratios as well as the mean difference between the noisy and noiseless transforms. 
For the inhomogeneous case, we follow the same procedure but for 2D images with added structures.

\bigskip
\subsubsection{Distribution of Radon Transform} \label{sec:ResultsDRadT}
\begin{figure}[h!]
    \centering
    \includegraphics[width=0.5\textwidth,trim={4.5cm 2cm 0cm 0cm},clip]{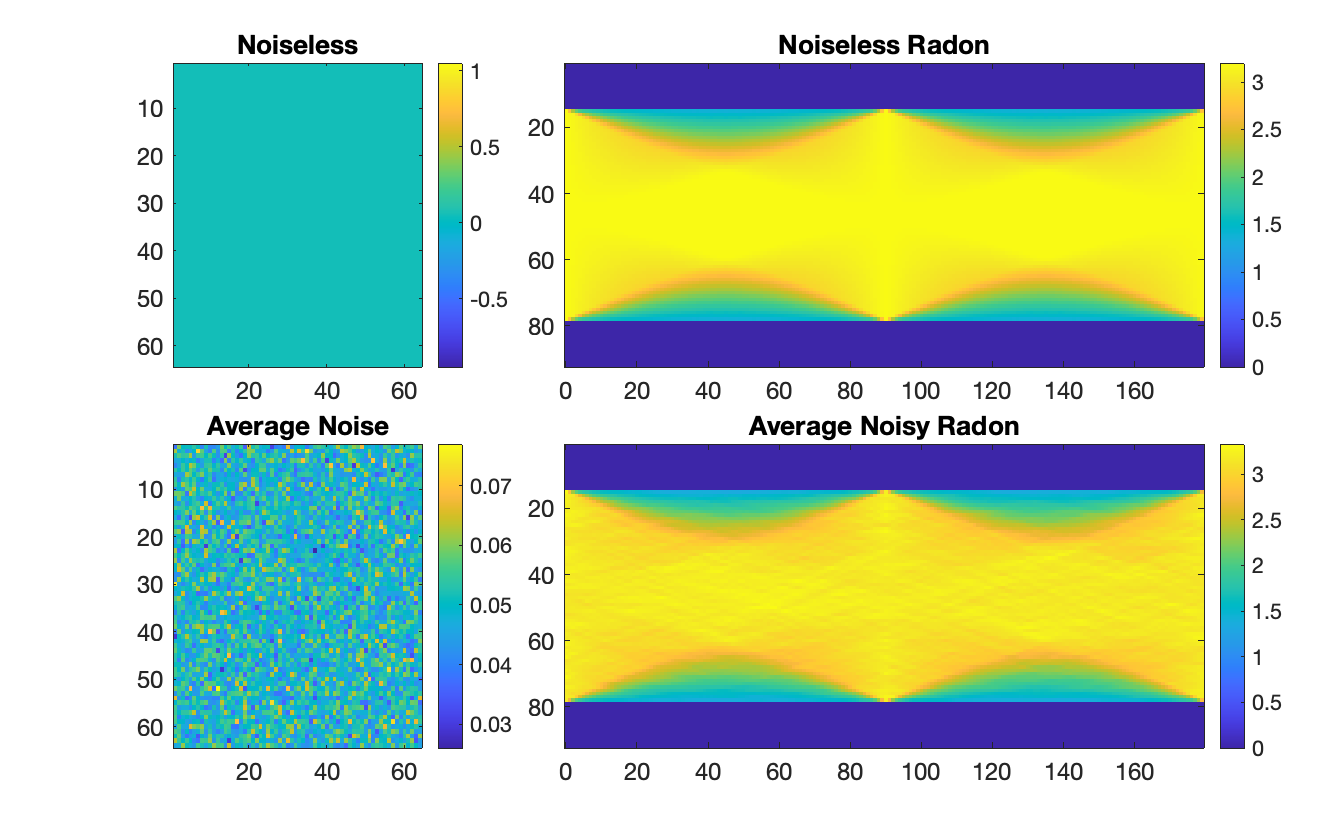}
    \caption{Visualisation of noiseless and average noisy Radon transforms of 1000 homgeneous Poisson-distributed images with intensity ${\lambda=0.05}$. } 
    \label{fig:radon_coeffs_homog}
\end{figure}

\begin{figure}[h!]
    \centering
    \includegraphics[width=0.5\textwidth,trim={4.5cm 2cm 0cm 0cm},clip]{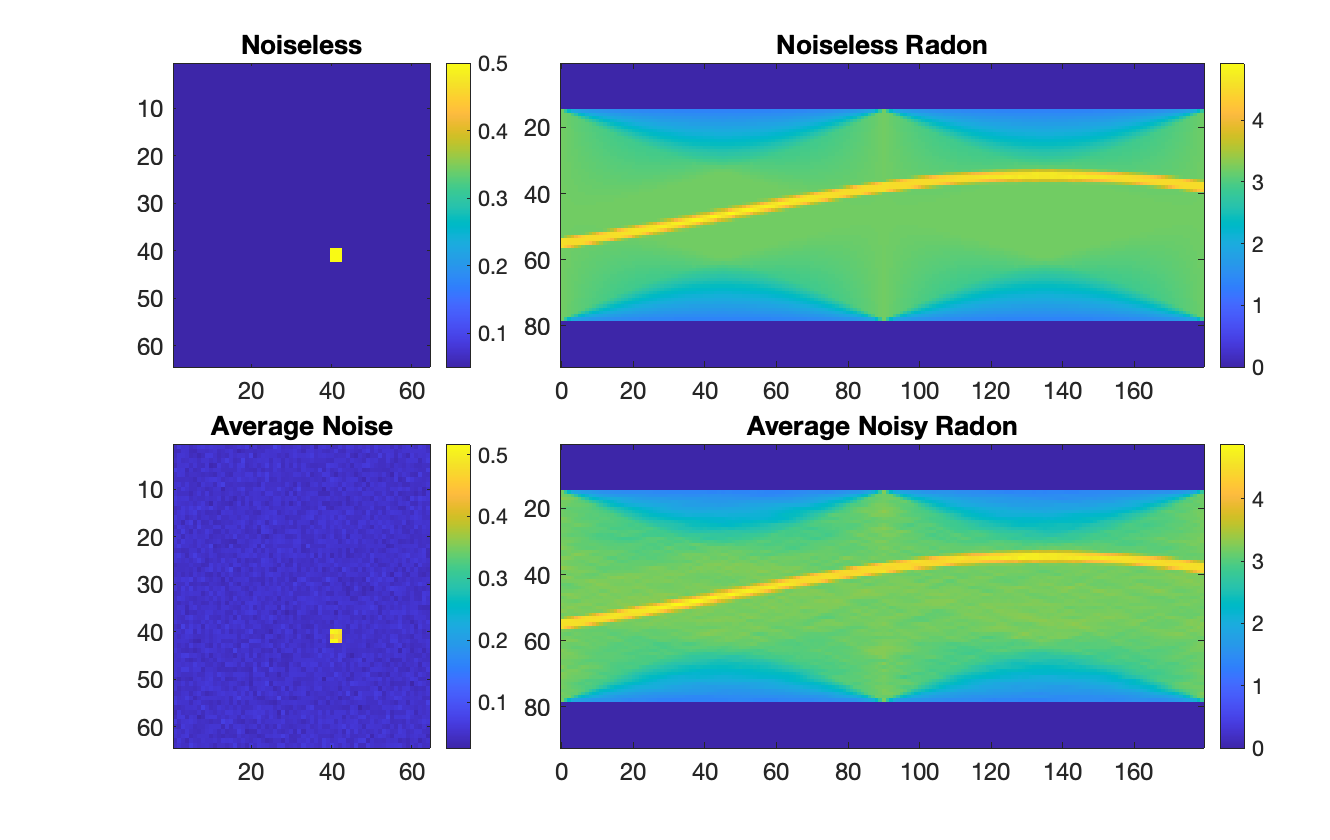}
    \caption{Visualisation of noiseless and average noisy Radon transforms of 1000 inhomgeneous Poisson-distributed images with intensity ${\lambda=0.05}$ and added structures of intentity $10\lambda$. 
    } 
    \label{fig:radon_coeffs_inhomog}
\end{figure}

\begin{figure}[h!]
    \centering
    \includegraphics[width=0.23\textwidth,trim={2.2cm 0.5cm 1.2cm 0cm},clip]{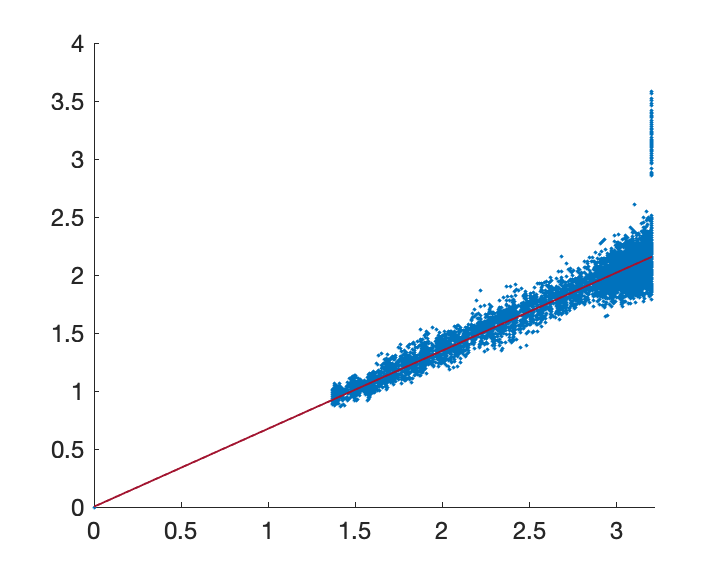}
    \includegraphics[width=0.23\textwidth,trim={2.2cm 0.5cm 1.2cm 0cm},clip]{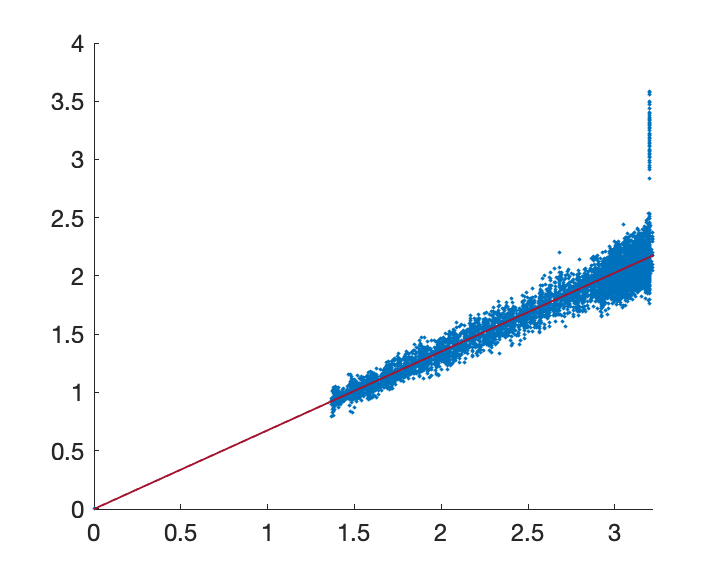}
    \caption{Noise analysis of the Poisson noise in the Radon domain showing noiseless pixel intensities verus noise variance for simulated Poisson-distributed images. Homogeneous case to the left, and inhomogeneous case to the right.  }
    \label{fig:radon_coeffs_scatters}
\end{figure}

This subsection reviews Poisson-noise generated images analysed in the Radon domain. Evaluations of the distributional properties of the noise were based on a distrete Radon transform implementing straightforward image rotations with $\theta\in[0,\pi)$. 
Figure \ref{fig:radon_coeffs_homog} and \ref{fig:radon_coeffs_inhomog} shows example images of the original and transforms noise in the homogeneous and inhomogeneous cases, respectively. The measured noise parameters shows that the transformed noise exhibits Poisson-characteristics such as equal mean and variance (mean variance ratios close to one); see Table \ref{tab:radon_coeffs}. Furthermore, Figure \ref{fig:radon_coeffs_scatters} indicates a linear relationship between the noiseless pixel intensity and the variance of the noise images. 
\subsubsection{Distribution of Ridgelet Transform} \label{sec:ResultsDRidgeT}
This subsection reviews Poisson-noise generated images analysed in the ridgelet domain. Evaluations of the distri-
\begin{table}[h!]
    \centering
    \footnotesize
    \caption{Estimated 95\% confidence intervals for noise parameters of the Radon coefficients based on 1000 noisy Poisson images with intensity $\lambda=0.05$. The inhomogeneous case includes additional structures of intentity $10\lambda$. \vspace*{0.1cm} }
    \def\arraystretch{1}
    \begin{tabular}{|lc|}
        \hline
        \multicolumn{2}{|l|}{\textit{Homogeneous case}} \\
        \hline
        pixel mean  & $( 1.950 , 1.971 )$ \\
        pixel variance & $( 1.312 , 1.326 )$ \\
        mean variance ratio & $( 1.033 , 1.044 )$ \\
        mean difference (noisy vs noiseless) & $( -0.005 , -0.004 )$ \\
        \hline
        \multicolumn{2}{|l|}{\textit{Inhomogeneous case}} \\
        \hline
        pixel mean & $( 1.994 , 2.016 )$ \\
        Pixel variance & $( 1.348 , 1.363 )$ \\
        mean variance ratio & $( 1.028 , 1.039 )$ \\
        mean difference (noisy vs noiseless) & $( -0.004 , -0.003 )$ \\
        \hline
    \end{tabular}
    \label{tab:radon_coeffs}
\end{table}

\noindent butional properties of the noise were based on a distrete Radon transform implementing straightforward image rotations with $\theta\in[0,\pi)$ followed by a wavelet transfrom of the Radon projections using Haar wavelets on one levels. For both homogeneous and inhomogeneous cases, we evalutate the distributional properties on approximation and detail coefficients separately. 

\begin{figure}[h!]
    \centering
    \includegraphics[width=0.5\textwidth,trim={4.5cm 2cm 0cm 0cm},clip]{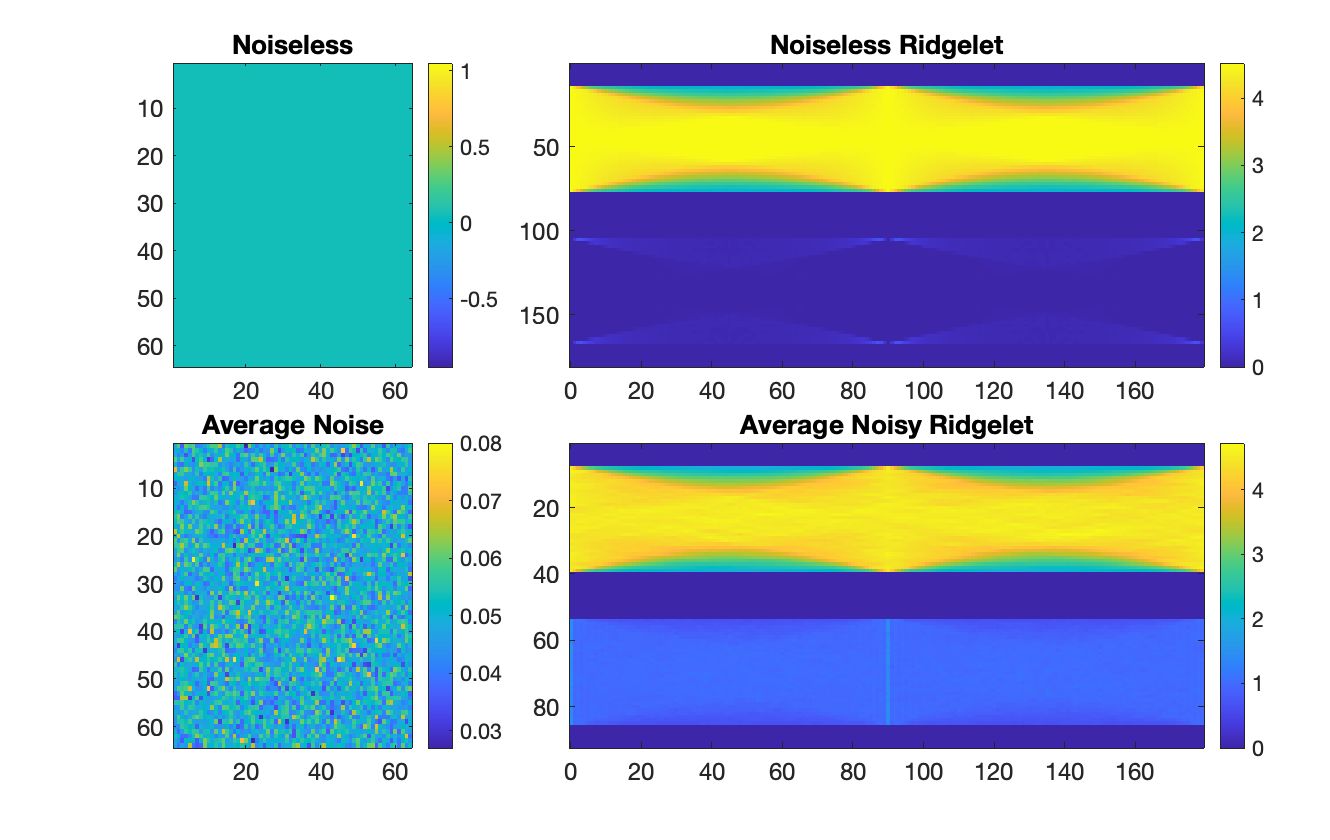}
    \caption{Visualisation of noiseless and average noisy ridgelet transforms based on 1-level Haar wavelets of 1000 homgeneous Poisson-distributed images with intensity ${\lambda=0.05}$. 
    }
    \label{fig:ridge_coeffs_homog}
\end{figure}

\begin{figure}[h!]
    \centering
    \includegraphics[width=0.5\textwidth,trim={4.5cm 2cm 0cm 0cm},clip]{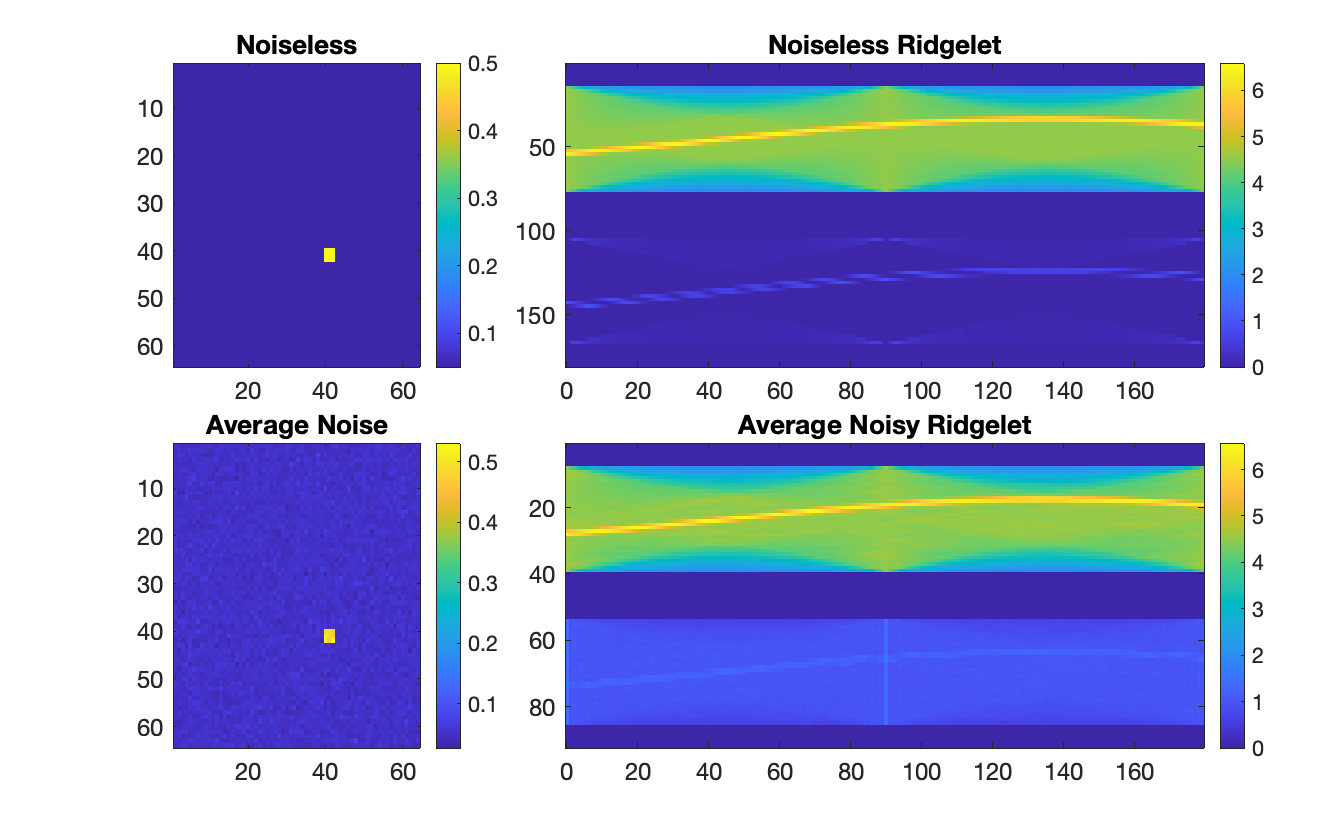}
    \caption{Visualisation of noiseless and average noisy ridgelet transforms based on 1-level Haar wavelets of 1000 inhomgeneous Poisson-distributed images with intensity ${\lambda=0.05}$ and added structures of intentity $10\lambda$. 
    }
    \label{fig:ridge_coeffs_inhomog}
\end{figure}

\begin{figure}[h!]
    \centering
    \includegraphics[width=0.23\textwidth,trim={2.2cm 0.5cm 1.2cm 0cm},clip]{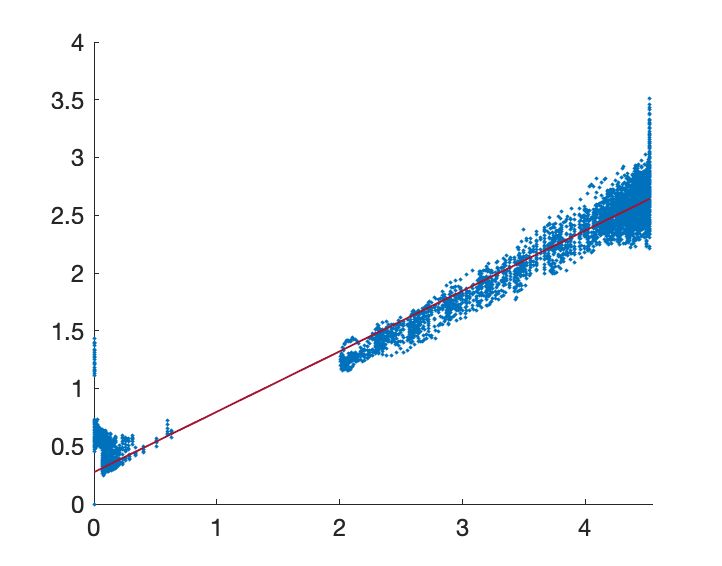}
    \includegraphics[width=0.23\textwidth,trim={2.2cm 0.5cm 1.2cm 0cm},clip]{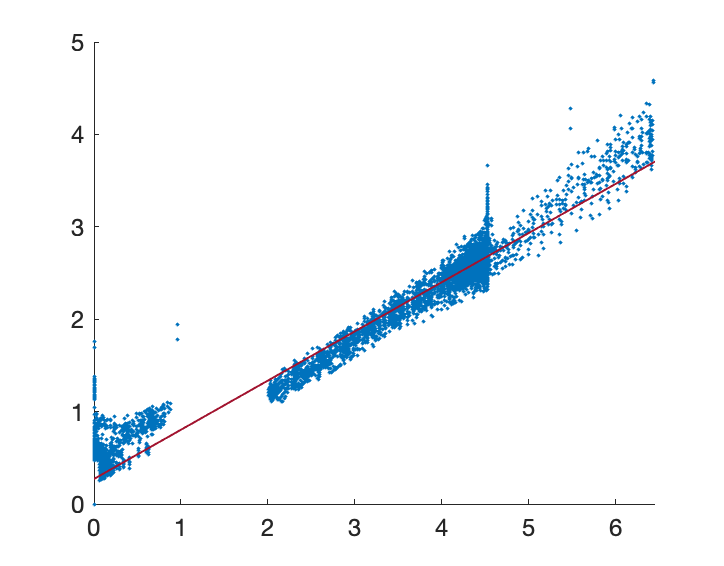}
    \caption{Noise analysis of the Poisson noise in the ridgelet domain based on 1-level Haar wavelets showing noiseless pixel intensities verus noise variance for simulated Poisson-distributed images. Homogeneous case to the left, and inhomogeneous case to the right.   }
    \label{fig:ridge_coeffs_scatters}
\end{figure}

Figure \ref{fig:ridge_coeffs_homog} and \ref{fig:ridge_coeffs_inhomog} shows example images of the original and transforms noise in the homogeneous and inhomogeneous cases, respectively. The measured noise parameters shows that the transformed noise exhibits Poisson-characteristics such as equal mean and variance (mean variance ratios close to one) in the detail layer coefficients; see Table \ref{tab:ridge_coeffs}. The proportional relationship between the noiseless pixel intensity and the variance of the noise images is shown in Figure \ref{fig:ridge_coeffs_scatters}. 

\begin{table}[h!]
    \centering
    \footnotesize
    \caption{Estimated 95\% confidence intervals for noise parameters of the Radon coefficients based on Poisson images with intensity $\lambda=0.05$. The inhomogeneous case includes additional structures of intentity $10\lambda$. \vspace*{0.1cm} }
    \begin{tabular}{|llcc|}       
        \hline
        \multicolumn{4}{|l|}{\textit{Homogeneous case}} \\
        \hline
        & pixel mean && $( 4.903 , 5.143 )$ \\
        \textit{Approximation} & pixel variance && $( 2.234 , 2.347  )$ \\
        \textit{coefficients} & mean variance ratio && $( 2.060 , 2.124 )$ \\
        & mean difference  && $(  0.002 , 0.009 )$ \\
        & (noisy vs noiseless) && \\
        \hline
        & pixel mean && $( 0.724 , 0.787 )$ \\
        \textit{Detail} & pixel variance && $( 0.439 , 0.478 )$ \\
        \textit{coefficients} &mean variance ratio && $( 1.179 , 1.278 )$ \\
        & mean difference && $( 0.639 , 0.698 )$ \\
        & (noisy vs noiseless) && \\
        \hline     
        \multicolumn{4}{|l|}{\textit{Inhomogeneous case}} \\
        \hline
        & pixel mean && $( 5.586 , 5.764 )$ \\
        \textit{Approximation} & pixel variance && $( 2.566 , 2.653 )$ \\
        \textit{coefficients} & mean variance ratio && $( 2.185 , 2.200 )$ \\
        & mean difference && $(  0.006 , 0.012 )$ \\
        & (noisy vs noiseless) && \\
        \hline
        & pixel mean && $( 0.715 , 0.780 )$ \\
        \textit{Detail} & pixel variance && $( 0.440 , 0.484 )$ \\
        \textit{coefficients} & mean variance ratio && $( 1.132 , 1.233 )$ \\
        & mean difference && $( 0.603 , 0.661 )$ \\
        & (noisy vs noiseless) && \\
        \hline
    \end{tabular}
    \label{tab:ridge_coeffs}
\end{table}


\bigskip
\subsection{Ridgelet Thresholding}
In this subsection, we illustrate the potential efficiency of Poisson noise removal using the ridgelet-based shrinkage estimator described in Section \ref{subseq:Thresholding using Stein method}. The estimator was implemented using Matlab's efficient built in frunction for Radon transform with $\theta\in[0,\pi)$, followed by wavelet transform of the Radon projections using 1-level undecimated Haar wavelets. Parameter estimation for the corresponding shrinkage rules were based on empirical risk minimisation. Evaluation is based on relative mean square errors (MSE), structural similarity index measures (SSIM), and peak signal-to-noise ratios (PSNR) for noisy and thresholded images. We define the evaluation measures as: 
\begin{align*}
MSE &= \frac{1}{n}\sum_{i=0}^n ( \hat{x}_i-x_i )^2  \\
SSIM &= l(\hat{x},x)^{\alpha} c(\hat{x},x)^{\beta} s(\hat{x},x)^{\gamma} \\
PSNR &= 10 \log \Big( \frac{x_{max}^2}{MSE} \Big)
\end{align*}
where $\hat{x}$ and $x$ are the noisy and noiseless images, respectively, with $n$ image pixels.
The SSIM components for luminance ($l$), contrast ($c$), and structure ($s$) are calculated based on means $\mu$, variance $\sigma^2$ and covariance $\sigma_{\hat{x}x}$ of the noisy and noiseless images as follows
\begin{align*}
l(\hat{x},x) &= \frac{2\mu_y\mu_{\hat{x}}+C_1}{\mu_x^2+\mu_{\hat{x}}^2+C_1} \\
c(\hat{x},x) &= \frac{2\sigma_y\sigma_{\hat{x}}+C_2}{\sigma_x^2+\sigma_{\hat{x}}^2+C_2} \\
s(\hat{x},x) &= \frac{\sigma_{x\hat{x}}+C_3}{\sigma_y\sigma_{\hat{x}}+C_3}
\end{align*}
where $C_1, C_2$ and $C_3$ are constants dependent of the dynamic range ($L$) of the images such that $C_1=(0.01L)^2$, $C_2=(0.03L)^2$ and $C_3=C_2/2$. For simplicity, we let $\alpha=\beta=\gamma=1$.

\begin{figure*}
    \centering
    \includegraphics[width=0.85\textwidth,trim={2cm 5cm 0cm 2cm},clip]{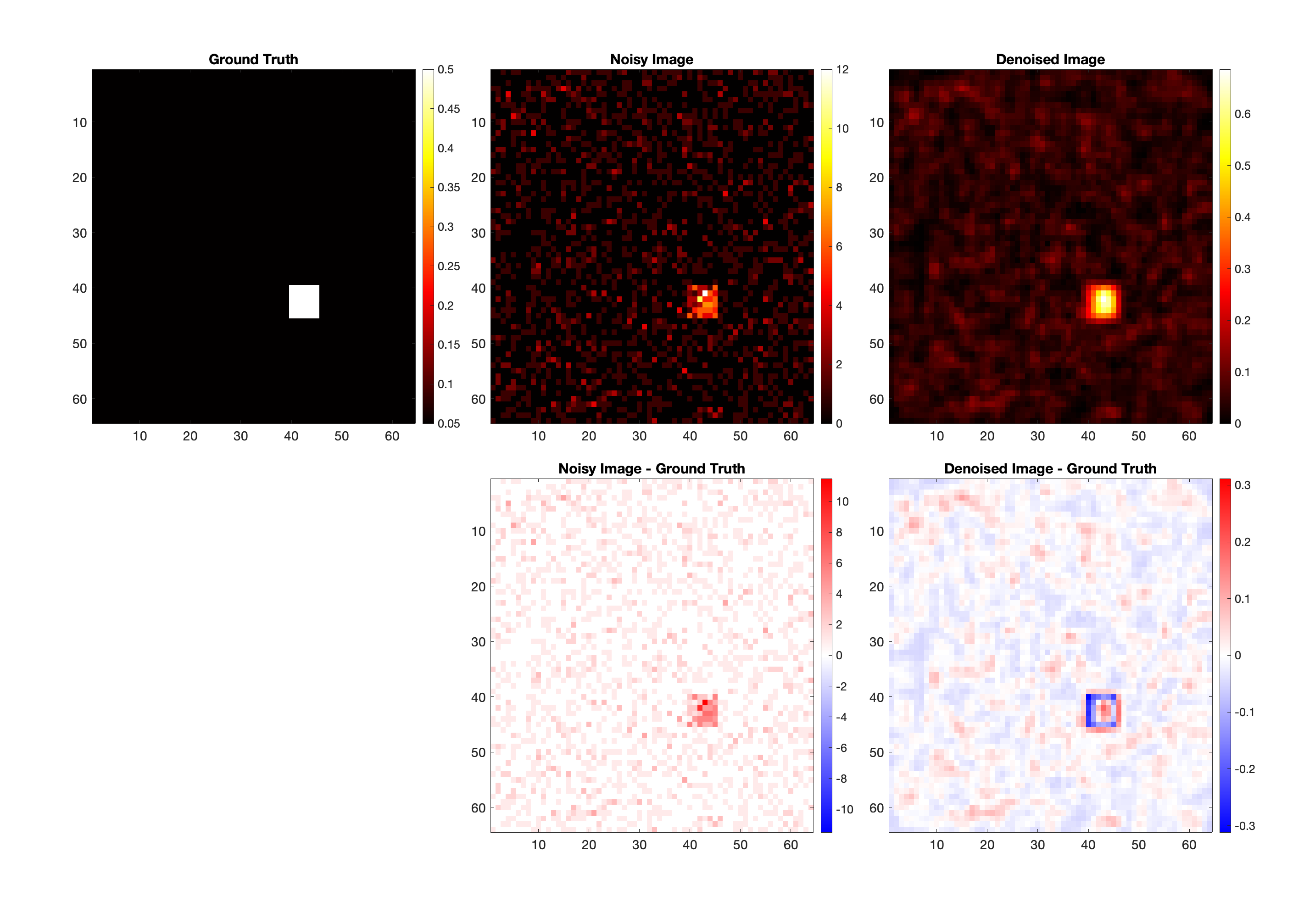}
    \caption{Thresholding experiments for Poisson-corrupted images showing the noiseless ground truth, the noisy image, and the denoised image based on ridgelet thresholding. Difference images betwwen noisy or denoised and ground truth are shown on the bottom row. }
    \label{fig:ridge_thresh}
\end{figure*}

We consider a simple 2D image with constatnt-intensity background and added structures following the inhomogeneous case in Section \ref{sec:Verification of Noise Distribution}. 
The resulting MSE, SSIM, and PSNR for 1000 noisy and densoised images are shown in Table \ref{tab:ridge_thresh}, and a visual comparison can be seen in Figure \ref{fig:ridge_thresh}.

\begin{table}[h!]
    \centering
    \footnotesize
    \caption{Estimated average (standard deviation) relative mean squared error (MSE), structural similarity index measure (SSIM), and peak signal-to-noise ratio (PSNR) based on 1000 samples of noisy and denoised data. \vspace*{0.1cm} }
    \begin{tabular}{|lcc|}       
        \hline
        & Noisy & Denoised \\
        \hline
        MSE & $0.820 \,( 0.039 ) $ & $ \, 8.383\cdot 10^4 \,( 4.434\cdot 10^5 )$ \\
        SSIM & $\qquad 0.002 \,( 1.289\cdot 10^4 ) $ & $  66.618 \,( 0.010 )$ \\
        PSNR & $49.00 \,( 0.204 ) $ & $  78.903 \,( 0.228 )$ \\
        \hline
    \end{tabular}
    \label{tab:ridge_thresh}
\end{table}

\begin{figure*}
    \centering
    \includegraphics[width=0.85\textwidth,trim={2cm 5cm 0cm 2cm},clip]{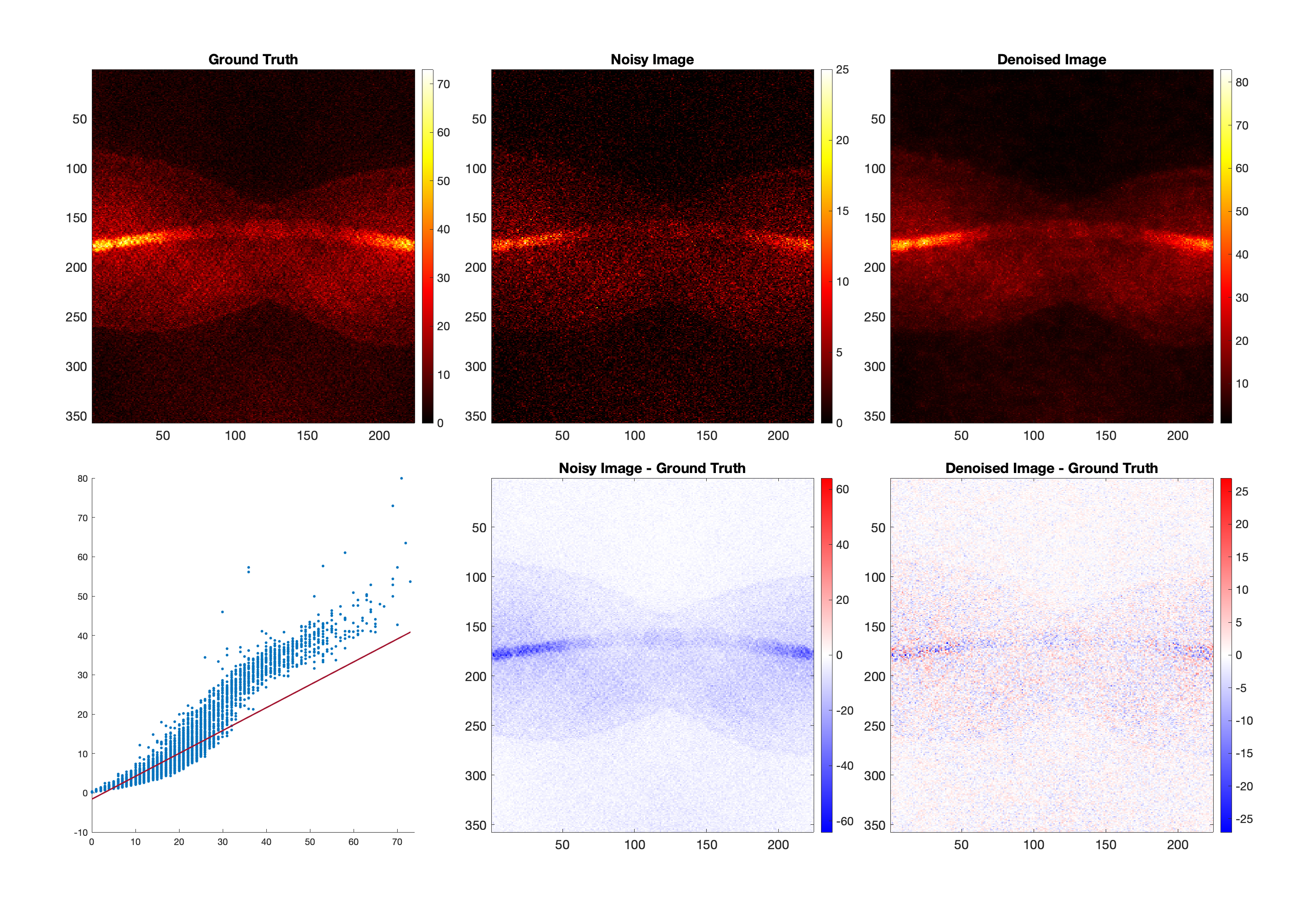}
    \caption{Thresholding experiments for Poisson-corrupted images showing the noiseless ground truth, the noisy image, and the denoised image based on ridgelet thresholding. Difference images betwwen noisy or denoised and ground truth are shown on the bottom row. 
    Noise analysis comparing noiseless pixel intensities verus noise variance for denoised images are shown at the bottom left. }
    \label{fig:ridge_threshPET}
\end{figure*}

To illustrate a potential area of application, we also consider ridgelet thresholding of Poisson data originating from PET measurements represented by 2D sinograms. The sinogram pixel values are considered as the true underlying intensity function of interest. We report on noise level characterisation and denoising efficiency in terms of MSE and SSIM. Implementations were set using undecimated Haar wavelets on 3 levels. 
The resulting MSE, SSIM, and PSNR for 1000 noisy and densoised images are shown in Table \ref{tab:ridge_threshPET}, as well as a visual comparison in Figure \ref{fig:ridge_threshPET}.

\begin{table}[h!]
    \centering
    \footnotesize
    \caption{Estimated average (standard deviation) relative mean squared error (MSE), structural similarity index measure (SSIM), and peak signal-to-noise ratio (PSNR) based on 1000 samples of noisy and denoised PET measurements. \vspace*{0.1cm}}
    \begin{tabular}{|lcc|}       
        \hline
        & Noisy & Denoised \\
        \hline
        MSE & $60.381 \,( 0.116 ) $ & $\,\,  7.655 \,( 0.055 )$ \\
        SSIM & $\,\, 0.146 \,( 0.001 ) $ & $\,\,  0.352 \,( 0.005 )$ \\
        PSNR & $30.322 \,( 0.008 ) $ & $  39.291 \,( 0.031 )$ \\
        \hline 
    \end{tabular}
    \label{tab:ridge_threshPET}
\end{table}


\balance
\bigskip
\section{Discussion}\label{Discussion}
In this study, we have investigated the distributional properties of noisy Poisson-distributed images in the Radon and wavelet domains in order to propose a proper thresholding scheme for noise removal in the ridgelet domain.  
We have shown, theoretically and experimentally, that the ridgelet transform preserves the Poisson distribution properties of the noise. This revelation is pivotal as it steers clear the pathway for leveraging the inherent characteristics of Poisson noise in the denoising process. Our theoretical framework proposes an adaptive thresholding approach in the ridgelet domain, guided by Stein's method. This approach not only aligns with the nature of Poisson noise but also opens up avenues for more refined denoising techniques.

The experimental part of our study brings to light several intriguing aspects.
Our noise analysis experiments revealed a tendency for underdispersion in the noise distribution. A plausible explanation for this could be the aggregated effect of pixel summation in the transforms, potentially elevating the mean more significantly than the variance. This might also imply a smoothing effect inherent in the summation process.
Furthermore, although transform coefficients in the radon domain indicate a linear noise model, this is not the case for ridgelet coefficients. Using wavelet decomposition on one level indicates that detail coefficients follow a linear noise model. 

The theoretical framework we have established lays the groundwork for developing a comprehensive denoising algorithm tailored for images corrupted with Poisson noise. The denoising results, as presented in our experiments, illustrate the potential of our proposed ridgelet thresholding technique. The immediate goal is to refine the ridgelet denoising technique, harnessing its potential to exploit Poisson distributional properties effectively.
Furthermore, the slight deviations observed in experimental patterns from our theoretical predictions highlight areas for future research. Delving deeper into these anomalies could unveil new insights into the behaviour of Poisson noise under various transforms and lead to even more robust denoising strategies.

In conclusion, our study represents a significant step in understanding and tackling Poisson noise in images through the lens of the ridgelet transform. The theoretical and experimental findings open up the development of advanced denoising techniques that are both effective and aligned with the intrinsic properties of Poisson noise. As we explore this domain, the horizon of possibilities in image processing and computer vision widens.

\section*{}
\vspace*{-0.5cm}
\subsection*{Funding}
This work is supported by the Swedish Research Council (340-2013-5342).
\vspace*{0.5cm}



\bibliography{bibliography}
\bibliographystyle{icml2023}
\end{document}

%% file: Our_macros.tex
\newcommand{\nsq}[1]{\|#1\|^2}

\newcommand{\R}{\mathbb{R}}

\newcommand{\T}{\mathbf{T}}

\definecolor{blue(munsell)}{rgb}{0.0, 0.5, 0.69}

\newcommand{\s}{\mathbf{s}}
\newcommand{\x}{\mathbf{x}}
\newcommand{\X}{\mathbf{X}}
\newcommand{\Xij}{X_{i, j}}

\newcommand{\XNji}{X_{N-1-j, i}}
\newcommand{\XiNj}{X_{i, N-1-j}}
\newcommand{\W}{\mathrm{W}}
\newcommand{\M}{\mathcal{M}_{a,b}(t)}

\newcommand{\Ra}{\mathrm{R}_{\X}}

\newcommand{\Raa}{\mathrm{R}_{\X}^{(1)}}
\newcommand{\Rab}{\mathrm{R}_{\X}^{(2)}}
\newcommand{\Rac}{\mathrm{R}_{\X}^{(3)}}
\newcommand{\Rad}{\mathrm{R}_{\X}^{(4)}}
\newcommand{\DN}{\mathcal{D}_N}
\newcommand{\DNN}{\mathcal{D}_{N/2}}

\newcommand{\Ri}{\mathcal{R}}

\newcommand{\lambdaij}{\lambda_{i, j}}
\newcommand{\lambdaji}{\lambda_{j, i}}
\newcommand{\lambdaNji}{\lambda_{N-1-j, i}}
\newcommand{\lambdaiNj}{\lambda_{i, N-1-j}}
\newcommand{\lambdahs}{\lambda_{h,s }}

\newcommand{\lambdabhs}{\lambda_{h,s }^{(2)}}
\newcommand{\lambdachs}{\lambda_{h,s }^{(3)}}
\newcommand{\lambdadhs}{\lambda_{h,s }^{(4)}}

\newcommand{\Sdist}{\mathcal{S}(\Tilde{\lambda}_{j,k}^+,\alpha_{j,k}^+,\Tilde{\lambda}_{j,k}^-, \alpha_{j,k}^-)}

\newcommand{\Qdist}{\mathcal{Q}(\Tilde{\lambda}_{j,k}^+,\alpha_{j,k}^+,\Tilde{\lambda}_{j,k}^-, \alpha_{j,k}^-)}

\newcommand{\cm}[1]{}